\documentclass{sig-alternate-10pt}

\usepackage{epstopdf}

\begin{document}

\title{Unified POF Programming for Diversified SDN Data Plane}

\numberofauthors{1}

\author{
\alignauthor Haoyu Song, Jun Gong, Hongfei Chen, Justin Dustzadeh\\
\affaddr {Huawei Technologies}}

\maketitle

\begin{abstract}

In many real-world OpenFlow-based SDN deployments, the ability to program heterogeneous forwarding elements built with different forwarding architectures is a desirable capability. In this paper, we discuss a data plane programming framework suitable for a flexible and protocol-oblivious data plane and show how OpenFlow can evolve to provide a generic interface for platform-independent programming and platform-specific compiling. We also show how an abstract instruction set can play a pivotal role to support different programming styles mapping to different forwarding chip architectures. As an example, we compare the compiler-mode and interpreter-mode implementations for an NPU-based forwarding element and conclude that the compiler-mode implementation can achieve a performance similar to that of a conventional non-SDN implementation. Built upon our protocol-oblivious forwarding (POF) vision, this work presents our continuous efforts to complete the ecosystem and pave the SDN evolving path. The programming framework could be considered as a proposal for the OpenFlow 2.0 standard.

\end{abstract}

\section{Introduction}

It has been envisioned that in SDN the network intelligence
should be moved to software as much as possible in order to support fast,
flexible, and low-cost network service deployments. Programmable forwarding
elements (FE) are essential to enable this vision.

While CPU- and NPU-based FEs are clearly
qualified candidates in term of programmability, they may
suffer a performance toll, especially in the scenario of data
center fabric where port density and aggregated bandwidth are
both very high. Therefore, at least by now they are more suitable
to be used in virtual switches at the edge or in routers in carrier
networks. On the other hand, the ASIC-based switch chips are
equipped with a fixed feature set but offer top port
density and throughput. Terabits throughput per chip is available today~\cite{trident2}. 
While not fully programmable,
these chips are configurable and able to handle
most of popular Data Center (DC) switch applications. 
ASIC-based FEs can be considered to have pre-installed 
packages or standard library functions. With
certain negotiation process such as TTP NDM~\cite{fawg}, ASIC-based
FEs can still be controlled under the
same SDN framework, as if they were programmed by the
controller. To truly fill the gap between performance and
programmability, a new breed of SDN-optimized chip is needed~\cite{stanfordTI}. 
With these chips, without compromising the performance,
network applications can be programmed on-the-fly and deployed in real time. 
Moreover, the system time-to-market is also reduced and the life cycle of FEs
extended.

For the foreseeable future, diverse FEs
built with different chips will coexist in various network segments.
As such, it is critical to have a unified framework, not only to
control and program these FEs, but also to hide the
heterogeneous substrate architecture and present a unified
programming interface to SDN controller and applications. We
envision OpenFlow to be the center pillar for this framework, however, 
further investigation and work are needed to address some of the challenges with the current approach, 
as articulated in~\cite{pof}. 

We believe the next generation of OpenFlow (e.g. OpenFlow 2.0) should offer the following
capabilities: (1) Allow the data plane to be protocol-oblivious so that no network behavior needs to be hardcoded
in FEs. This capability is important to ensure SDN
extensibility and programmability. (2) Allow the SDN controller to
be agnostic to FE architecture so that the data plane abstraction can help isolate the
controller from the FE implementation details.
This capability is important to allow SDN to sustain the heterogeneous
substrate platforms while still enjoying the programming freedom.
(3) Allow coexistence of coarse-grained programming through the
use of packages or library functions and fine-grained
programming through the use of flow instructions. This powerful capability  
extends the usability of diversified FEs and can offer the needed 
flexibility to satisfy most, if not all of the requirements of SDN users and developers.
While these goals may appear audacious, we believe they
represent the right direction for the evolution of OpenFlow and are
achievable with the right architecture and design decisions. In this paper, we
present an OpenFlow-based SDN programming framework and 
provide our experience on realizing it.

\section{Unified Programming Framework}

\begin{figure}[!ht] \centering
\includegraphics[width=\columnwidth]{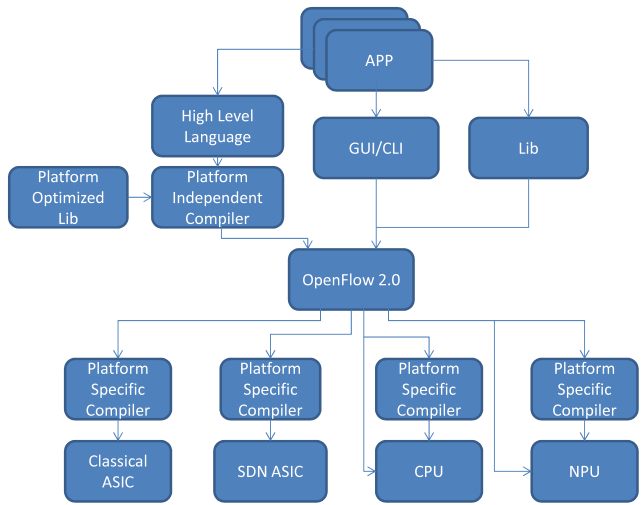} \caption{Proposed SDN
Data Plane Programming Framework} \label{fig_framework}
\end{figure}

The unified data plane programming framework is depicted in
Figure~\ref{fig_framework}. The center pillar of this framework
is the OpenFlow interface which provides a set of generic
instructions as well as other dataplane provision mechanisms.
This is the part that needs to be standardized. It provides a
decoupling point between the control plane and the data plane.
Ideally, and as discussed in~\cite{pof}, this interface should have 
versatile, future-proof, and
protocol/platform-agnostic properties.

\subsection{Intermediate OpenFlow Interface}

OpenFlow is pivotal for the data plane programmability. 
The key to a successful design of such a programming interface is to make it work at the
right abstraction levels. In particular, the interface should not be tied to a particular
FE architecture. Instead, it should offer the ability to be easily mapped to
any implementation while allowing for some specific optimizations to fully exploit the FE capability, if desired.

The core of our proposed OpenFlow interface is a set of generic ``flow"
instructions. These instructions function as the intermediate
vehicle between the platform-independent programming
environment and each individual target platform. The instructions
are grouped and summarized as follows:

\begin{itemize}

\item \emph{Packet/Metadata editing:} set field, add field,
delete field, math/logic operations on field

\item \emph{Flow Metadata manipulation:} read, write \item
\emph{Algorithm/Function procedure:} checksum, fragmentation,
etc.

\item \emph{Table access:} go to table (non return), search
table (return to calling instruction)

\item \emph{Output:} physical/virtual/logical port,
sampling/data-path generated packets

\item \emph{Jump/Branching:} conditional and unconditional,
absolute and relative

\item \emph{Active data path:} insert/delete/modify flow entry,
insert/delete flow table
  
\end{itemize} 

In addition to making the flow instructions protocol-oblivious, we 
propose other new features to enhance the programmability
and to enable performance optimization. 
One notable addition is the ability to abstract the instructions and actions associated with each
flow entry as a piece of program. This provides several
advantages. For example, it allows decoupling match keys and
actions. The actions for flow entries, in form of instruction
blocks, can be downloaded to FEs separately from flow entry
installations. When each instruction block is assigned a unique
ID, the flow entry only needs to include a block ID to infer
the associated actions. By doing this, not only different flow entries
can share the same instruction block while an instruction
block is only downloaded and stored once, but also there would 
theoretically be no limit on how many instructions one flow entry
can execute. The current OpenFlow flow model does not offer such a
capability. Our proposed model also allows easy instruction block updates:
one can simply load a new instruction block, update the block ID
in affected flow entries, and then revoke the old instruction
block if it is not needed anymore.

To facilitate instruction block sharing and at the same
time enable differentiated flow treatment, we propose to augment the
flow entry with a parameter field. This field can be leveraged
by application developers to define any parameters used by the associated instruction block. For example, in an
egress table, when all the entries execute an output action,
they may have different target output ports. While the output
action is coded in an instruction block and shared by all the
flow entries, the output port number is stored in the parameter
field of each flow entry. This is just an overly
simplified example. In reality, this mechanism is very powerful
to reduce the code space and complexity.

We also abstract the globally-shared memory resource as a flow
metadata pool. Flow metadata can be shared by flow entries to store
statistics (i.e. counters) or any other information such as
flow states. This is another enhancement on top of the existing
packet metadata mechanism which is only dedicated to each
packet. In particular, the expressivity of flow metadata enables stateful
dataplane programming.

\subsection{Programming over OpenFlow Interface}

Above the OpenFlow interface, any network forwarding application needs to
be converted to the standard OpenFlow instructions first. There are
three ways to do it. First, it would be easy to provide some high-level language to program network applications. The high-level
language provides another layer of abstraction that supports
modularity and composition~\cite{sdnlang}. With the help of a high-level
language, developers can focus on what
the application really wants to achieve rather than dealing with particular FE
architecture and conducting tedious and error-prone flow-level
match-action manipulations. Quite a few such languages have been proposed
in literature~\cite{frenetic, nettle, ppipp}. Since many modern
chips are C-programmable~\cite{ezchip, netronome} and C language is well-known
and widely used, we are exploring the possibility of using C as
our choice of high level language. However, this is still an
open and active research area. Until we thoroughly fathom the
feasibility, we do not exclude other possibilities.

Although programming in a high-level language is meant to be
forwarding-platform-independent, we realize that in the near
future, many different forwarding architectures will coexist.
For example, some chips (notably ASIC-based chips) have a
front-end packet parser which parses packets in a
centralized way but some other chips (notably NPU-based chips)
have a distributed packet parser which parses the packets layer
by layer along the packet processing pipeline. Moreover, each
kind of chip may have its own feature extensions, hardware-accelerated
modules, and other nuances in hardware resource provisioning.
Without discerning these differences, a generic program would
pose significant challenges to the complier which may lead to
poor performance or even worse, failure to compile at all.
Therefore, the application program should follow some
programming style upfront and may include some preprocessor
directives to guide the compiling process. The key point is that the language itself must be
general enough. The platform-independent compiler
compiles the application programs by calling the platform-optimized library. This is not a perfect solution from a
purist's perspective. However, as long as the FE
chips do not converge to a single architecture, we have to live
with it.
  
Another method is to directly use GUI/CLI for interactive data
plane programming. This could be considered similar to programming in assembly
language. Although it needs to handle flow level details, this
method is fast and direct. The GUI/CLI can be used to handle fast updates and can also be used to directly
download compiled applications to data plane FEs. 
We have implemented an open-source GUI to support
this programming method~\cite{pofweb}.
  
At last, there are many prevailing network applications and
forwarding processes today. For example, the basic L2
switching and L3 IP forwarding are still widely used. It would
be counterproductive to try to develop them again and again.
Also, some applications on some particular target platforms may
have been deeply optimized to achieve the best possible
performance. It would be very difficult for inexperienced
developers to implement these applications with a similar
performance. Therefore, pre-compiled applications can be
provided in a library by any third party and directly used to
program the network. Conceptually, this is in line with the
Table Type Pattern (TTP) developed by ONF FAWG~\cite{fawg}. Once
the specifications of these library applications are
standardized or publicized, any third party can develop and
release them. Users can also maintain their private library and
download the program through GUI or CLI.

Note that these programming approaches are not mutually exclusive. In other words, 
an application could be implemented through the simultaneous use of 
more than one approach. In a typical scenario, the basic forwarding
process is either customized by using the high-level language or taken 
from a standard library application, and then GUI/CLI
is used for library application download, dynamic runtime
updates, and interactive monitoring.

\subsection{Programming Diversified Platforms}

Each type of FEs may have its own
platform-dependent compiler which compiles the programs in standard OpenFlow
instructions to its local structures. We roughly categorize
FEs into four groups based on the type of main
forwarding chips on them.

\subsubsection{Conventional ASIC-based}

Conventional ASICs for FEs typically have a
fixed feature set and are not openly programmable. However, since they
are designed to handle classical forwarding scenarios at high
performance, they are still usable in SDN but in a more
restrictive way. In this case, the standard library applications
are the most suitable way to program the FEs.
Some ASICs are configurable and can switch between different
modes to support different applications. In this case, 
customized programming is not impossible but needs to be 
applied in a highly-disciplined way to ensure compatibility.

\subsubsection{SDN ASIC-based}

Recent research has started to pay more attention to
SDN-optimized chips~\cite{hardware, stanfordTI}. Many companies
are planing or have started to develop chips to better support
flexible network application programming~\cite{xpliant,
netronome}. These chips have embedded
programmable capability for general packet handling but are also
heavily populated with hardware-accelerated modules to handle
common network functions for high performance. For these
chips, it is feasible to use any kind of programming method. A
compiler is needed to compile the standard OpenFlow instructions
to the chip's local structure.

A compiler, no matter how well-designed, may cause some
performance loss due to the extra level of indirection. When the
OpenFlow 2.0 is standardized, it is conceivable that in the
future we could even design a chip that can natively execute the
OpenFlow instructions without even needing a compiler in data
plane.

\subsubsection{CPU-based}

CPU is no doubt the most flexible platform. Albeit having lower
performance compared with the other platforms, it can easily
support any programming method. Software-based virtual switches
are widely used in data centers. The switch
implementation in CPU can basically run in two different modes:
compiler mode and interpreter mode. 
The former compiles an application (in the intermediate form of OpenFlow instructions) 
into machine binary code and the latter requires the forwarding plane to directly interpret and execute OpenFlow instructions.
The interpreter mode is more
straightforward to implement. The open source soft switch in~\cite{pofweb} works in interpreter mode. It is
unclear to us which mode has higher
performance. We are working on a compiler-mode
implementation based on x86 platform which targets on virtual switches. 

\subsubsection{NPU-based}

Network Processing Units (NPU) are software programmable chips
that are designed specifically for network applications. An NPU
typically contains multiple processing cores to enhance the
parallel processing capability. NPUs can be broadly categorized
into two types: pipeline and run-to-completion (RTC).  

A representative pipeline NPU is EZchip's NP family chip~\cite{np}. In a pipeline NPU, each stage processor only handles a portion of packet
processing tasks. Although the pipeline NPU's architecture seems to match OpenFlow's processing pipeline model, in reality it is not easy to perfectly map the two pipelines together because OpenFlow's pipeline is function-oriented and NPU's pipeline is performance-oriented. The compiler needs to carefully craft the job partition to balance the load of pipeline stages. 

In an RTC NPU, each processor core is
responsible for the entire processing of a packet. This
architecture maximizes the programming flexibility which is similar to CPUs. However, it has
limited code space per core and needs to share resources (e.g.
memory) among cores. The code space constraint requires the code size to be compact enough in order to accommodate the whole processing procedure (e.g. we cannot afford to repeat the storage of the same set of actions for every flow in a large flow table). The resource sharing constraint requires both the number of memory accesses and the transaction size per memory access to be minimized in order to meet the performance target. Fortunately, the new features we proposed for the OpenFlow 2.0 interface allow software developers to program efficiently with these constraints in mind.

NPU-based FEs can also be programmed in compiler mode or interpreter mode. In the next section, we discuss the implementations of both modes on an NPU-based FE and compare their performance.

\section{NPU-based Case Study}

The NPU-based FE prototype works on Huawei's NE-5000
core router platform. The line card we used has an in-house
designed 40G NPU and each half slot interface card has eight
1GbE optical interfaces. The multi-core NPU runs in RTC mode. 

\subsection{Forwarding Programming in C}

To support high level data plane programming, we model three entities: Metadata, Table, and Packet. The program
simply manipulates these three entities and forwards the
resulting packets. For our NPU, the three entities are all realized in registers. Metadata is used to hold the packet metadata which is represented as a customized structure; Table is the associated data of flow entries loaded from table matches, which
is also represented as a customized structure; Packet is
typically the packet header under process which is described in
another structure.

The following example shows the structures of Metadata, Table,
and Packet for an L3 forwarding application:

{\small
\begin{verbatim}
struct Metadata_L3 {
    uint8		L3Stake;	//L3 Offset
    uint16		VpnID;	//VPN ID
    uint16		RealLength;	//Packet Length
    uint16		SqID;		//QOS Queue ID
};
struct Table_Portinfo {
    uint16		VpnID;	//VPN ID
    uint16		SqID;		//QOS Queue ID
};
struct IPV4_HEADER_S {
    uint4		Version;			
    uint4		HeaderLength;
    union {
        uint8   TOS;                
        uint6   DSCP;               
        uint3   Precedence;         
    };
    uint16		TotalLength;                 
    uint16		FragReAssemID;                     
    IPV4_FRAG_HWORD_S		FragHWord;
    IPV4_TTL_PROT_HWORD_S		TtlProtWord;
    uint16  Checksum;
    uint32  SIP;                   
    uint32  DIP;                
};
\end{verbatim}
}

A piece of program that processes a packet is shown below. It combines the IP address and the VPN ID as a new key to conducts another table lookup. 

{\small
\begin{verbatim}
(Metadata_L3 *) p_metadata;
(Table_Portinfo *) p_table;
p_metatada->VpnID = p_table->VpnID;
p_ipheader = p_packet + 14;
Goto_Table(TableID, p_metadata->VpnID, p_ipheader->DIP);
\end{verbatim}
}

Once the packet processing flow is described in C, it is straightforward to compile the program into intermediate OpenFlow instructions. Although the programming style appears to be platform independent, the Goto\_Table library function could be specific for each different forwarding platform. To infer the different platform implementation to the compiler, an NPU-specifc proprietary library is included.

\subsection{Interpreter Mode FE Implementation}

In interpreter mode, each intermediate OpenFlow instruction corresponds to a piece of code written in NPU microcode which realizes the instruction's function. The code translation is straightforward. However, due to the flexibility embedded in the OpenFlow instructions, the efficiency of the microcode is problematic. 

For example, the Goto\_Table instruction may lead to a complex microcode processing flow. First, it needs to read the corresponding table information and initialize a buffer to hold the search key, then it enters a loop to construct the search key piece by piece depending on the number of header fields involved in the instruction. Each iteration of the loop contains many steps. It needs to locate the target field using the  offset and length information, copy the field into the key buffer, and mask the field. This process requires a lot of pointer shift, data move, and other logic operations. Finally, the search key is sent to the target flow table and the thread is hung up to wait for the lookup result.   

The inefficiency comes from three sources: (a) the microcode instruction count, (b) the number of thread switch, and (c) the bandwidth of loading flow table entries. The microcode instruction count is determined by the microcode instruction set and the complexity of the OpenFlow instructions. The thread switch is caused by the loops that force to break processing pipelines as well as the latency for table lookups. Each table lookup will return an instruction block. If parameters are directly carried within instructions, the bandwidth of loading such instruction blocks are considerably expanded. As a result, the throughput suffers.  

\subsection{Compiler Mode FE Implementation}

In compiler mode, the compiling process can significantly simplify the microcode.
Since there are a set of registers $R0 \sim Rn$ in NPU, the compiler can resolve the pointer offsets and directly map the data into registers. This eliminates the need of pointer manipulations in microcode. The compiler also handles the length evaluation and directly translates that into assignment statement. These can help to reduce the microcode instruction count by more than 50\%.   

The compiler mode implementation takes advantages of the flow parameter mechanism which significantly reduces the instruction block size. This lowers the bandwidth requirement for memory access and further boosts the throughput and latency performance.

\subsection{Performance Evaluation}

The packet forwarding performance in NPU is evaluated by throughput ($R$) and packet latency ($L$). We know that  
$R= c*f/i$ and $L = t/R$ in which $c$ is the number of processing cores, $f$ is core frequency, $i$ is microcode instruction count per core, and $t$ is the number of threads. Given an NPU, $c$ and $f$ are fixed, so the performance is mainly determined by $i$ and $t$. Reducing table lookup latency and memory access bandwidth have direct impact on $t$. Table~\ref{tb_gototable} compares the performance of different Goto\_Table implementations ($n$ is the number of match fields in the search key). 

\begin{table}[!ht]
\centering 
\begin{tabular}{|l|l|l|}\hline
 & instr. count & \# thread switch\\\hline\hline
 Interpreter Mode  & $37+33n$ & $7+3n$ \\\hline
 Compiler Mode  & 13+n &  1\\\hline
\end{tabular}
\caption{\label{tb_gototable} \emph{Goto\_Table} Performance Comparison}
\end{table}

Table~\ref{tb_eval} summarizes the performance comparison for basic IPv4 forwarding. The conventional non-SDN implementation is used as a benchmark. The conventional implementation can fully take advantage of the hardware features and the microcode is deeply optimized.    

\begin{table}[!ht]
\centering 
\begin{tabular}{|l|l|l|l|}\hline
  & non-SDN & Interpreter & Compiler \\\hline\hline
 instr. count & 496 & 1089 & 550  \\\hline
 \# thread switch & 94 & 146 & 74\\\hline
 thruput (Mpps) & 77.5 & 35.3 & 69.8 \\\hline
 latency (cycle) & 4468 & 6361 & 4022 \\\hline
\end{tabular}
\caption{\label{tb_eval} Performance comparison for Basic \emph{IPv4} Forwarding}
\end{table}

Through extensive experiments, we found that the compiler-mode
implementation performs consistently better than the
interpreter-mode implementation. For a typical IP forwarding
process in routers, the compiler-mode implementation needs 57\%
less microcode instructions than the interpreter-mode
implementation. Compared with the conventional 
implementation, the compiler-mode implementation is just 11\%
worse. With the same number of micro cores, a compiler-mode
implementation can easily double the throughput of an
interpreter-mode implementation.

\section{Related Work}

P4 describes an abstract forwarding model as a
strawman proposal for OpenFlow 2.0~\cite{ppipp}. It uses the
platform-independent language $P4$ to define the header
parse graph and the switch control program. The control
program basically describes the table types and the action set
supported by each table. The model also needs a platform-dependent
compiler to map the configuration to each specific target
switch. After configuration, the controller can then populate
the tables with flow entries at run time. This architecture
allows flexible parsing and editing too. However, the model is more 
restrictive than ours in programmability because (1) the action set for each table
need to be predefined in the ``configure" phase; (2) it only supports a front-end parser which can be problematic
for some specific applications. The model does not clearly show where the OpenFlow
interface should be located. If the platform-dependent
configuration compiler is located in switch, then the $P4$
configuration would appear on the OpenFlow interface, and in turn the
specifications of parse graph, action set, and control program
need to be standardized.

OCP networking project advocates open switches with
open-programming environments~\cite{ocp}. Quite a few open
switch specifications and open-source softwares have been
released since the project debut in 2013. However, at its current stage this project still falls short of SDN support:
(1) It focuses on programming in an open Linux-based NOS environment for each individual switch
but not in a centralized SDN programming environment; (2) The
current open switch specifications heavily rely on existing
ASIC-based chips and SDK/API provided by chip
vendors. The programming flexibility is limited by the chip
architecture and the degree of openness the chip vendors would
like to offer. We believe a truly open switch also means
open silicon chips or at least a universal and complete API. 
The project might evolve towards a
similar direction as we proposed.

\section{Conclusions}

We believe it is plausible to assume that the next generation SDN will require 
total programmability over an open data plane. An FE could be programmed as easily as a server can be programmed today.
However, the
diversified chips used to build the FEs today and in the foreseeable future are far from a 
convergence. This poses a serious
challenge for the desired uniform and coherent SDN programming experience. 
Until we solve this problem,
we cannot claim a vertical-decoupling of the SDN layered architecture is fully achieved. 
With the current SDN approach, it could become very difficult to build an efficient   
ecosystem in which players would work at
different layers independently. 

In this paper we present our initial exploration and experience
on this hard problem. We propose a possible programming
framework which centers on the next-generation OpenFlow
interface, targets various FEs, and supports
different programming approaches. In particular, we experiment
on an NPU-based platform and show that the complier-mode implementation
is superior to the interpreter-mode implementation. Apart from
other factors, the microcode instruction count plays an
important role in determining the throughput performance.
Compiler mode provides excellent match between the OpenFlow
instructions and the microcode instructions. In light of this,
the ultimate performance can be gained by one-on-one direct
instruction mapping and the capability of native OpenFlow instruction execution
in FE chips. This is in the domain of research for future
SDN-specific chips. OpenFlow 2.0 designers need to
work with chip vendors closely to consider this possibility.

Our future work includes completing the proposed SDN programming
framework by implementing the missing pieces in
Figure~\ref{fig_framework} (e.g. platform-dependent
compilers for other FE platforms) and demonstrating real-world SDN
applications through the full programming process. This
programming framework can be considered as a proposal for the
OpenFlow 2.0 standard.

{\small
\bibliographystyle{IEEEtran} 
\bibliography{sigproc}

\begin{thebibliography}{10}
\providecommand{\url}[1]{#1}
\csname url@samestyle\endcsname
\providecommand{\newblock}{\relax}
\providecommand{\bibinfo}[2]{#2}
\providecommand{\BIBentrySTDinterwordspacing}{\spaceskip=0pt\relax}
\providecommand{\BIBentryALTinterwordstretchfactor}{4}
\providecommand{\BIBentryALTinterwordspacing}{\spaceskip=\fontdimen2\font plus
\BIBentryALTinterwordstretchfactor\fontdimen3\font minus
  \fontdimen4\font\relax}
\providecommand{\BIBforeignlanguage}[2]{{%
\expandafter\ifx\csname l@#1\endcsname\relax
\typeout{** WARNING: IEEEtran.bst: No hyphenation pattern has been}%
\typeout{** loaded for the language `#1'. Using the pattern for}%
\typeout{** the default language instead.}%
\else
\language=\csname l@#1\endcsname
\fi
#2}}
\providecommand{\BIBdecl}{\relax}
\BIBdecl

\bibitem{trident2}
\BIBentryALTinterwordspacing
(2013) {Trident II Switch}. [Online]. Available: \url{http://www.broadcom.com/}
\BIBentrySTDinterwordspacing

\bibitem{fawg}
\BIBentryALTinterwordspacing
(2013) {ONF} {F}orwarding {A}bstraction {W}orking {G}roup ({FAWG}). [Online].
  Available:
  \url{https://www.opennetworking.org/working-groups/forwarding-abstractions}
\BIBentrySTDinterwordspacing

\bibitem{stanfordTI}
P.~Bosshart, G.~Gibb, H.-S. Kim, G.~Varghese, N.~McKeown, M.~Izzard, F.~Mujica,
  and M.~Horowitz, ``{Forwarding Metamorphosis: Fast Programmable Match-action
  Processing in Hardware for SDN},'' in \emph{{Proceedings of the ACM
  SIGCOMM}}, 2013.

\bibitem{pof}
H.~Song, ``{P}rotocol-{O}blivious {F}orwarding: {U}nleash the {P}ower of {SDN}
  through a {F}uture-{P}roof {F}orwarding {P}lane,'' in \emph{{ACM SIGCOMM
  HotSDN Workshop}}, 2013.

\bibitem{sdnlang}
N.~Foster, M.~Freedman, A.~Guha, R.~Harrison, N.~P. Katta, C.~Monsanto,
  J.~Reich, M.~Reitblatt, J.~Rexford, C.~Schlesinger, A.~Story, and D.~Walker,
  ``{Languages for Software Defined Networks},'' \emph{{IEEE Communication
  Magazine}}, Feburary 2013.

\bibitem{frenetic}
N.~Foster, R.~Harrison, M.~J. Freedman, C.~Monsanto, J.~Rexford, A.~Story, and
  D.~Walker, ``{F}renetic: {A} {N}etwork {P}rogramming {L}anguage,'' in
  \emph{{ACM SIGPLAN ICFP}}, 2011.

\bibitem{nettle}
A.~Voellmy and P.~Hudak, ``{N}ettle:{F}unctional {R}eactive {P}rogramming of
  {O}pen{F}low {N}etworks,'' in \emph{PADL}, 2011.

\bibitem{ppipp}
P.~Bosshart, D.~Daly, M.~Izzard, N.~McKeown, J.~Rexford, D.~Talayco, A.~Vahdat,
  G.~Varghese, and D.~Walker, ``{Programming Protocol Independent Packet
  Processors},'' in \emph{{Unpublished}}, 2013.

\bibitem{ezchip}
\BIBentryALTinterwordspacing
(2012) {EZ}chip {NPS}. [Online]. Available: \url{http://www.ezchip.com/}
\BIBentrySTDinterwordspacing

\bibitem{netronome}
\BIBentryALTinterwordspacing
(2012) {N}etronome {F}low {P}rocessor. [Online]. Available:
  \url{http://www.netronome.com/}
\BIBentrySTDinterwordspacing

\bibitem{pofweb}
\BIBentryALTinterwordspacing
(2013) {P}rotocol {O}blivious {F}orwarding. [Online]. Available:
  \url{http://www.poforwarding.org}
\BIBentrySTDinterwordspacing

\bibitem{hardware}
M.~Casado, T.~Koponen, D.~Moon, and S.~Shenker, ``{R}ethinking {P}acket
  {F}orwarding {H}ardware,'' in \emph{{ACM SIGCOMM HotNets Workshop}}, November
  2008.

\bibitem{xpliant}
\BIBentryALTinterwordspacing
(2013) {X}pliant. [Online]. Available: \url{http://www.xpliant.com/}
\BIBentrySTDinterwordspacing

\bibitem{np}
{Ran Giladi}, \emph{{Network Processors: Architecture, Programming, and
  Implementation (Systems on Silicon)}}.\hskip 1em plus 0.5em minus 0.4em\relax
  {Morgan Kaufmann}, 2008.

\bibitem{ocp}
\BIBentryALTinterwordspacing
(2013) {Open Compute Project}. [Online]. Available:
  \url{http://www.opencompute.org/}
\BIBentrySTDinterwordspacing

\end{thebibliography}
}
\end{document}